Chapter 11

# Increasing cognitive-emotional flexibility with meditation and hypnosis
## The cognitive neuroscience of de-automatization

Kieran C. R. Fox, Yoona Kang, Michael Lifshitz, and Kalina Christoff

## Abstract

Meditation and hypnosis both aim to facilitate cognitive-emotional flexibility, i.e., the "de-automatization" of thought and behavior. However, little research or theory has addressed how internal thought patterns might change after such interventions, even though alterations in the internal flow of consciousness may precede externally observable changes in behavior. This chapter outlines three mechanisms by which meditation or hypnosis might alter or reduce automatic associations and elaborations of spontaneous thought: by an overall reduction of the chaining of thoughts into an associative stream; by de-automatizing and diversifying the content of thought chains (i.e., increasing thought flexibility or variety); and, finally, by re-automatizing chains of thought along desired or valued paths (i.e., forming new, voluntarily chosen mental habits). The authors discuss behavioral and cognitive neuroscientific evidence demonstrating the influence of hypnosis and meditation on internal cognition and highlight the putative neurobiological basis, as well as potential benefits, of these forms of de-automatization.

## Introduction

When left free and unconstrained, what do people think about of their own accord? Or rather, since the process of thought seems to be largely spontaneous and uncontrollable, what does the mind-brain think about of *its* own accord? Is there any possibility of increasing the flexibility and diversity of these spontaneous chains of thought? Although cognitive psychologists have been addressing this question for several decades now (e.g., Klinger, 2008; Singer, 1966; Singer & McCraven, 1961), cognitive and clinical neuroscientists are only just beginning to seriously apply scientific scrutiny to such "spontaneous" thought processes (Andrews-Hanna, Smallwood, & Spreng, 2014; Fox & Christoff, 2014, 2015; Fox, Spreng, Ellamil, Andrews-Hanna, & Christoff, 2015; Smallwood & Andrews-Hanna, 2013).

Spontaneous thought processes have been studied under various names, including daydreaming, mind wandering, and stimulus-independent thought, although they are not always characterized by "spontaneity" per se. (For a discussion of terminology, see Christoff, 2012; Dixon, Fox, & Christoff, 2014; Fox & Christoff, 2014.) For the sake of simplicity, we use these terms more or less interchangeably in this chapter, wherein we focus not on nuances of terminology but, rather, on outlining what is known about the subjective content and neural basis of these spontaneous thought processes. Although the subjective *content* of spontaneous thought has been fairly well studied over the past few decades (e.g., Klinger, 2008; Singer, 1966; Singer & McCraven, 1961), and is now being addressed with large-scale studies utilizing experience sampling with thousands of participants (e.g., Killingsworth & Gilbert, 2010), we still know comparatively little about the cognitive *processes* underlying it. What seems intuitively apparent, however, is that most of our spontaneous thought is characterized by *automaticity*—our streams of thought flow quickly, without conscious effort, and along habitual, well-worn paths.

## Spontaneous thought: phenomenological content, neural correlates, and cognitive automaticity

In this introductory section, we outline three main features central to an understanding of spontaneous thought processes: the *phenomenological content*, as reported by first-person reports from many studies over the past few decades; the preliminary understanding we have of the *neural correlates* of various spontaneous thought processes; and, finally, the evidence concerning the *cognitive automaticity* of most of our streams of thought. We then turn our attention to a number of ways that our automatized patterns of thought may be made more flexible and diverse, and present arguments for why such de-automatization might be beneficial.

### Phenomenological content and neural correlates

First-person accounts reveal spontaneous thought to be a highly varied and complex phenomenon, drawing on all sensory modalities, reaching into the distant past and anticipated future, and spanning the intellectual gamut from escapist fantasy to scientific and artistic creativity (Andrews-Hanna, Reidler, Huang, & Buckner, 2010; Fox, Nijeboer, Solomonova, Domhoff, & Christoff, 2013; Fox, Thompson, Andrews-Hanna, & Christoff, 2014; Klinger, 1990, 2008, 2013; McMillan, Kaufman, & Singer, 2013; Smallwood & Andrews-Hanna, 2013). The cognitive neuroscientific study of spontaneous thought lags far behind this phenomenological work, but a general picture of brain activity contributing to the various forms of "mind wandering" is beginning to emerge, implicating a correspondingly wide range of brain regions (see Fox et al., 2015; see also Figure 11.1 and Plate 4).

The salient involvement of the medial prefrontal cortex, posterior cingulate cortex, and a swathe of lateral parietal lobe (including the temporoparietal junction and parts of the inferior parietal lobule) strongly parallels the core regions of the so-called "default mode" network of brain regions that are consistently recruited during the resting state (Buckner, Andrews-Hanna, & Schacter, 2008). The medial temporal lobe, critical to memory formation and recall, as well as prospection (imagining the future) (Addis, Pan, Vu, Laiser, & Schacter, 2009; Schacter, Addis, & Buckner, 2007), is also consistently recruited (Fox et al., 2015). Finally, there is strong evidence for recruitment of the lateral prefrontal cortex, dorsal anterior cingulate cortex, temporopolar cortex, and the insula during mind wandering—findings often neglected in both empirical and theoretical work on spontaneous thought (Fox et al., 2015).

Although the specific functional contributions and temporal dynamics of these various brain areas still remain poorly understood (Andrews-Hanna et al., 2014; Fox et al., 2015),

AQ1

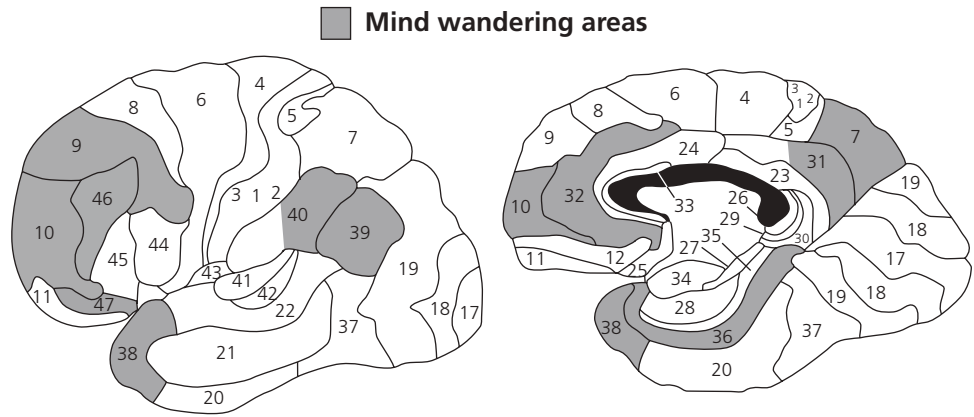

**Fig. 11.1** Brain regions consistently recruited by various forms of spontaneous thought (see also Plate 4).

Approximate Brodmann areas (BAs) consistently recruited in functional neuroimaging studies of various spontaneous thought processes. Note the contribution of medial temporal lobe structures that may be involved in generating memories and imagined future scenarios (BA 36, 38), as well as lateral prefrontal cortex areas potentially underlying goal-directed planning and selection among various spontaneous thoughts.

Adapted from K. Brodmann, *Vergleichende Lokalisationslehre der Großhirnrinde in ihren Prinzipien dargestellt auf Grund des Zellenbaues*, Barth, Leipzig, Copyright © 1909.

some tentative steps have been made toward an understanding of these interrelationships. One preliminary conception put forth by our group divides spontaneous thought into roughly three stages (see Table 11.1), which preferentially (though not exclusively) recruit specific brain areas implicated in spontaneous thought in general (see Figures 11.1 and 11.2 and Plate 4). In the first *generative* stage, spontaneous thoughts arise their own accord, via mechanisms as yet unknown. Medial temporal lobe structures, particularly the hippocampus and parahippocampus, seem likely sources of many of the memories and imagined future scenarios that form the basis for much of spontaneous conceptual thought (Addis et al., 2009; Ellamil, Dobson, Beeman, & Christoff, 2012; Ellamil et al., in preparation; Gelbard-Sagiv, Mukamel, Harel, Malach, & Fried, 2008), whereas the insula may contribute to higher-order, spontaneous "thoughts" about the visceral state of the body (Craig, 2004, 2009; Critchley, Wiens, Rotshtein, Öhman, & Dolan, 2004).

In the subsequent *elaborative* stage, initially isolated thoughts are associated to other thoughts, memories, and emotions, and may develop into a *stream* of thought that tends to follow habitual paths and tendencies. The medial prefrontal cortex (PFC), with its strong involvement in self-referential processing (D'Argembeau et al., 2007; Northoff et al., 2006), and the posterior cingulate cortex and temporopolar cortex, which are strongly tied to recall, elaboration, and processing of memory (Andrews-Hanna, Reidler, Sepulcre, Poulin, & Buckner, 2010; Christoff, 2013; Christoff, Ream, & Gabrieli, 2004; Svoboda, McKinnon, & Levine, 2006), seem likely to be involved in this elaborative process (Ellamil et al., in preparation; Farb et al., 2007).

A final *evaluative* or *executive* stage may sometimes occur, in which thoughts are monitored, directed, and possibly selected for their relevance to the self and long-term goals. The most likely players at this stage appear to be the dorsal anterior cingulate cortex, as well as the dorsolateral and

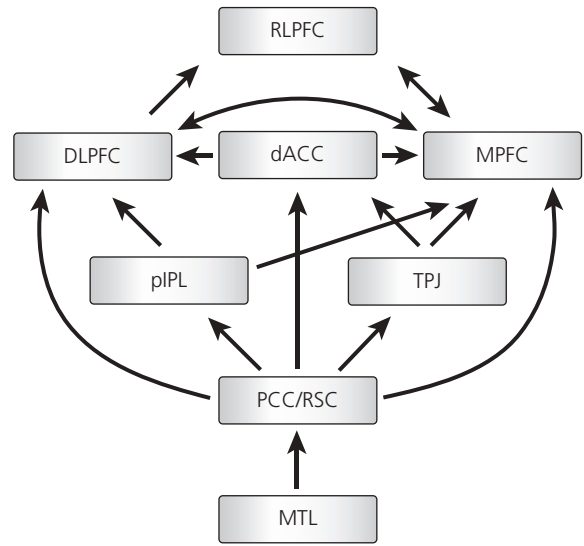

**Fig. 11.2** A preliminary model of neural pathways underlying internally generated spontaneous thought.

A preliminary model suggests that many spontaneous thoughts may, at least in part,originate in the medial temporal lobe (MTL; especially hippocampus and parahippocampus) as either spontaneously recurring memories or imagined future scenarios (the *generation* stage; see Table 11.1), and are then elaborated upon by numerous other brain regions such as the posterior cingulate cortex (PCC; BA 31) and/or retrosplenial cortex (RSC), and medial prefrontal cortex (MPFC) in an *elaboration* stage. Finally, "higher" areas such as the dorsolateral prefrontal cortex (DLPFC; BA 9/46), rostrolateral prefrontal cortex (RLPFC; BA 10), and dorsal anterior cingulate cortex (dACC) may be recruited to evaluate, select among, and guide these streams of thought (the *evaluation* stage). Note that this model does not include occurrence of spontaneous thoughts related to the body's various interoceptive signals, or thoughts based directly on incoming stimuli from the external world; see Dixon et al. (2014) for such details. Unidirectional arrows are meant to indicate the primary direction of information flow, but all interregional connections should be considered reciprocal. pIPL: posterior inferior parietal lobule (BA 39/40); TPJ: temporoparietal junction (BA 39/40).

Adapted from *Neuropsychologia*, 62, Matthew L. Dixon, Kieran C.R. Fox, and Kalina Christoff, A framework for understanding the relationship between externally and internally directed cognition, pp. 321–30, doi:10.1016/j.neuropsychologia.20140.05.024, Copyright 2014, Elsevier. With permission from Elsevier.

rostrolateral prefrontal cortex—regions often recruited during spontaneous thought, metacognition, and top-down cognitive control (see Figure 11.1) (Andrews-Hanna et al., 2014; Christoff, Gordon, Smallwood, Smith, & Schooler, 2009; Fox & Christoff, 2014; Fox et al., 2015).

## Cognitive automaticity

Still less studied than the phenomenological content and neural basis of spontaneous thoughts, is the degree to which they chain together into *streams* of thought, and the tendency of these streams to follow predictable, habitual paths—that is, the degree to which spontaneous thinking is characterized by "automaticity" (see the section "The automaticity of spontaneous thought: advantages and disadvantages"). Automaticity might characterize any or all stages of spontaneous thought (see Table 11.1 and Figure 11.2): we might predictably generate ideas of a certain nature; elaborate upon such ideas in a habitual fashion; or evaluate our own spontaneous thoughts and emotions in habitual, characteristic judgments and valuations.

**Table 11.1** Three possible stages of spontaneous thought

| Stage | Core contributing brain regions* |
|---|---|
| **Generation**<br>Origination/creation of new thoughts and imagery; spontaneous recall of memories, or imagination of future scenarios; recombination of experiences in memory consolidation and reconsolidation | Medial temporal lobe (hippocampus, parahippocampus); posterior cingulate cortex; temporoparietal junction |
| **Elaboration**<br>Spontaneously arising thought is spun out into a stream of associated thoughts and emotions | Medial prefrontal cortex; temporopolar cortex |
| **Evaluation**<br>Spontaneous thoughts or streams of thought are judged and evaluated for their personal utility or emotional valence; possible guidance/steering of streams of thought by metacognitive brain areas | Lateral prefrontal cortex; dorsal anterior cingulate cortex |

*Brain regions listed appear to show relatively greater contributions at a given stage, but the lists should by no means be considered definitive or mutually exclusive. See also Figure 11.2 for a graphical illustration of the regions involved and a preliminary model of the functional neuroanatomical flow of spontaneously arising thoughts through various brain areas.

   When might such automaticity be helpful—and when harmful? Are there ways in which we can increase the flexibility and diversity of the cognitive and emotional aspects of our spontaneous thought? Practices geared toward changing both the content and process of spontaneous thought are as ancient as their scientific study is new—such techniques have existed for millennia and continue to be developed in modern clinical contexts. In this chapter, we discuss two such methods and highlight the cognitive neuroscientific evidence supporting their effectiveness: the ancient practices of meditation, and the comparably recent methods of clinical (i.e., "hypnotherapy") and experimental hypnosis.

   Meditation and hypnotherapy both begin with the simple premise that patterns of everyday thought and behavior are suboptimal for virtually everyone, and can be improved to increase one's well-being. The problem is not only that we fail to live up to particular moral paragons endorsed by given people, societies, or religions; more importantly, we often even fail to think the way *we* want to think, feel the way we want to feel, and do the things we want to do. It seems that, in many ways, we are not free to choose our own actions, much less our own thoughts and feelings: a large, perhaps dominant, portion of our thinking and behavior is guided instead by habitual thought patterns, automatic behaviors, and default emotional reactions (which we collectively refer to here using the umbrella term "automaticity"—see the section "The automaticity of spontaneous thought: advantages and disadvantages"). On the milder end of the spectrum, this can lead to regrets, missed opportunities, biased opinions, and unfair treatment of ourselves and others; in its more pernicious forms, to self-destructive behaviors such as drug and alcohol addiction, clinical syndromes such as depression, anxiety, or post-traumatic stress disorder, and widespread societal issues such as racism and sexism. Both meditation and hypnotherapy, however, appear to offer systematic methods of de-automatizing these habitual thought patterns.

# The automaticity of spontaneous thought: advantages and disadvantages

It seems then that many of our spontaneous thought processes are supported by heuristic routines that are processed automatically, outside of conscious awareness (Bargh & Chartrand, 1999). We use the term *automaticity* to refer to the process of effortlessly engaging in behaviors, or patterns of thinking, according to previously established associations, without conscious monitoring (cf. LaBerge & Samuels, 1974). Automaticity is usually a desired result of learning that reflects mastery and fluency, and can help lessen the self-regulatory burden by freeing up limited cognitive resources from tasks for which they are no longer needed (e.g., Bargh & Chartrand, 1999). Compared to deliberate cognitive processes, automatic processes are fast, relatively effortless, and tax few cognitive resources.

Although automaticity can therefore be a beneficial aspect of cognition, and is indeed a necessary part of life, automatized cognitive or emotional reactivity can also potentially lead to a wide range of detrimental outcomes. At the individual level, automatic reactivity to events may subserve maladaptive thought patterns prevalent in mental health disorders, such as lack of perceived control in anxiety disorders (Chorpita & Barlow, 1998), addiction (Forsyth, Parker, & Finlay, 2003), and negative rumination in depression (Nolen-Hoeksema, 1991). At the societal level, an often highly automatic and unconscious process of intergroup attitude formation and maintenance (Monteith, Zuwerink, & Devine, 1994) can activate negative and stereotypic associations, unconsciously influencing one's opinions of, and behavior toward, outgroup members of another race or economic class (Bargh, 1989; Fiske, Cuddy, & Glick, 2007; Gaertner & Dovidio, 2008).

Early efforts to de-automatize maladaptive mental habits were made in the field of clinical psychology. In the landmark analysis of depression, for instance, Beck (1967) characterized the disorder as involving automatic chains of irrational thoughts and established a therapeutic technique based on de-automatization. His attention-based cognitive therapy involves bringing attention to the largely automatic sequence of irrational thought patterns in order to break (i.e., de-automatize) them. The long-term aim is the replacement of old maladaptive mental habits with new, more adaptive ones (which we refer to as *re-automatization*; see the section "Re-automatization of spontaneous thought processes"). More recent findings further suggest that de-automatization of resistant mental habits is also possible in the social domain; for example, counter-stereotyping egalitarian goals can be preemptively activated to eliminate stereotyping tendencies (Moskowitz & Li, 2011).

Many of these studies and interventions aim only for short-term effects, however—a "proof of concept" that thought and behavior can be modified by interventions, rather than a large-scale overhaul of mental habits. There appear to be more thorough and systematic forms of de-automatization though that might prove more effective over the long term—namely, meditation training and hypnotic suggestion. The remainder of this chapter will explore various ways in which automatized patterns of thought and emotional reactivity can be made more flexible and diverse via these methods. We draw primarily on functional neuroimaging research that has begun to shed some light on the possible neural mechanisms underlying de-automatization, and appears to support the efficacy of meditation and hypnosis in altering these processes.

# Recognizing and combating the automaticity of spontaneous thought processes: the benefits of meta-awareness

Although some unconscious methods of de-automatization have already been alluded to (e.g., Dasgupta & Greenwald, 2001; Moskowitz & Li, 2011; Sassenberg & Moskowitz, 2005), the first step in the conscious and systematic de-automatization of spontaneous thought processes is considered to be the *recognition* of thought's automaticity in the first place. Few of us even recognize how often we are mind wandering (Christoff et al., 2009; Fox & Christoff, 2015; Schooler et al., 2011), much less the true degree to which our thoughts and emotions simply follow well-trod paths. The experience of psychiatrist and neuroscientist Roger Walsh (1977), when he first began practicing meditation, exemplifies this realization:

> I was forced to recognize that what I had formerly believed to be my rational mind preoccupied with cognition, planning, problem solving, etc., actually comprised a frantic torrent of forceful, demanding, loud, and often unrelated thoughts and fantasies which filled an unbelievable proportion of consciousness even during purposive behavior. The incredible proportion of consciousness which this fantasy world occupied, my powerlessness to remove it for more than a few seconds, and my former state of mindlessness or ignorance of its existence, staggered me . . . Foremost among the implicit beliefs of orthodox Western psychology is the assumption that man spends most of his time reasoning and problem solving, and that only neurotics and other abnormals spend much time, outside of leisure, in fantasy. However, it is my impression that prolonged self-observation will show that at most times we are living almost in a dream world in which we skillfully and automatically yet unknowingly blend inputs from reality and fantasy in accordance with our needs and defenses . . . The subtlety, complexity, infinite range and number, and entrapping power of the fantasies which the mind creates seem impossible to comprehend, to differentiate from reality while in them, and even more so to describe to one who has not experienced them. (Walsh, 1977, p. 154)

Despite the pervasiveness of such automaticity, it is encouraging that a frank recognition of the power and frequency of automatic thought processes does not require years of self-observation: Walsh's (1977) insight is typically among the first realizations experienced by a beginning practitioner of meditation (Gunaratana, 2011). In some of our recent work (Fox & Christoff, 2014; Kang, Gruber, & Gray, 2013), we have discussed how metacognitive awareness of such habitual patterns of mind wandering might be a key element in breaking down these persistent chains of thinking, and re-orienting them towards desired and valued alternative patterns of thought. For instance, a recent fMRI (functional magnetic resonance imaging) study from our group compared brain activity during mind wandering both with, and in the absence of, meta-awareness of the fact that one was mind wandering (Christoff et al., 2009). We found that although very similar brain regions were activated in both cases (compared to periods when subjects were not mind wandering), numerous regions were significantly *less* active when meta-awareness was present than when it was absent (see Figure 11.3 and Plate 5; see also Table 3 in Christoff et al., 2009). When meta-awareness was present, significantly less brain activation was observed in numerous medial temporal lobe regions strongly implicated in long-term memory (Schacter et al., 2007) and the generation of spontaneous thoughts (Christoff et al., 2004; Fox et al., 2015); in downstream visual brain areas, such as the fusiform and lingual gyri, implicated in mental imagery and the imagined visual scenes accompanying dreaming (Domhoff & Fox, 2015; Fox et al., 2013; Solms, 1997); and in the medial prefrontal cortex, implicated in self-referential processing and the elaboration of streams of thought (see Figure 11.3). Recently, we have shown that these regions (among others) become increasingly active when one moves from externally directed thought, to daydreaming, to full-blown and immersive dream mentation (see Figure 3 in Fox et al., 2013; see also Domhoff and Fox, 2015).

One interpretation of these results is that meta-awareness during mind wandering dampens the *generative* and *elaborative* stages of spontaneous thought, either by reducing the

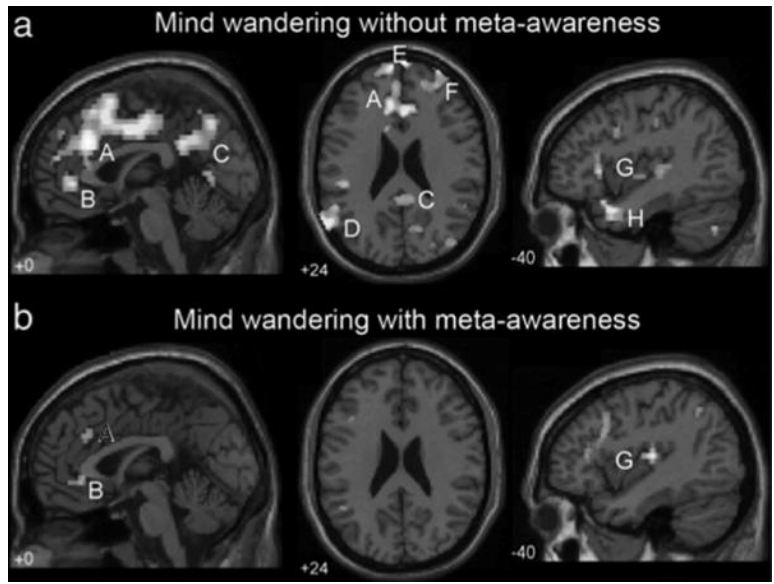

**Fig. 11.3** Differences in brain activity between mind wandering with and without meta-awareness (see also Plate 5).

Although mind wandering both with (panel b) and without (panel a) meta-awareness recruits a similar set of brain regions, recruitment is much more widespread for the latter (panel a). This pattern may suggest that meta-awareness attenuates, or otherwise exerts selective pressure upon, the stream of spontaneous thought (Fox & Christoff, 2014).

quantity of thoughts by selectively choosing which thoughts are allowed to take shape (Fox & Christoff, 2014) or, perhaps, by preventing such deep immersion in these streams of thought (Fox et al., 2013). (Both notions are suggestive of a "non-elaborative" mental stance; see the section "Reducing the chaining of spontaneous thoughts: a non-elaborative mental stance"). The notion that meta-awareness is allowing only selective elaboration of certain streams of thought, or otherwise attenuating the intensity of daydreaming (Fox & Christoff, 2014; Schooler et al., 2011), is supported by behavioral studies which show that performance on externally directed tasks is significantly worse when meta-awareness of mind wandering is absent than when it is present (Smallwood, McSpadden, Luus, & Schooler, 2008; Smallwood, McSpadden, & Schooler, 2007). These performance deficits may therefore be due to greater elaboration of thoughts in the absence of meta-awareness, and/or a deeper immersion in streams of thought (see also the section "Reducing the chaining of spontaneous thoughts: a non-elaborative mental stance").

Once the occurrence (i.e., generation) of automatic spontaneous thought (Figure 11.4a) is recognized, there are several different ways in which the habitual course of the stream of thought might be changed (see Figure 11.4 and Table 11.2). Perhaps the simplest change (at least conceptually, but certainly not in practice) is a reduction of the chaining together of thoughts into associative streams (Figure 11.4b), by adopting what is sometimes referred to as a "non-elaborative" mental stance. Another possibility is the general "de-automatization" of thought patterns (Figure 11.4c), such that a given thought becomes likely to lead to a number of further thoughts—instead of always chaining together habitually with a particular subsequent idea. *Re*-automatization of spontaneous thought (Figure 11.4d), on the other hand, involves the voluntary reformation of thought associations along new, specific habitual paths.

**Table 11.2** Overview of automatized and more flexible patterns of spontaneous thought

| Cognitive process | Overview |
|---|---|
| Automatized spontaneous thought | The usual form of thought for most people, most of the time: thoughts are deeply immersive and reified, and chain together in habitual patterns, centering on particular topics. Original, creative thought and new patterns of emotional reactivity are relatively rare. |
| Meta-awareness/non-elaborative mental stance | Conscious awareness and/or monitoring of spontaneously arising thoughts either dampens the intensity and immersive nature of thoughts, or prevents them from chaining together into associative streams of thought beyond one's control. |
| De-automatized spontaneous thought | Thoughts are not necessarily less immersive or less liable to chain together into associative streams, but the pattern of these streams is now less rigid and habitual: the stream of thought is broader, more flexible, more likely to follow novel paths. |
| Re-automatized spontaneous thought | Patterns of thought are deliberately reformed along new paths that are "automatized" but intentionally chosen. Desirable or personally useful patterns of thinking that are at first difficult and resource-draining are cultivated to become automatic mental habits. |

In the following sections, we discuss each of these possibilities in turn. By these divisions, we do not in any way mean to imply that these processes necessarily, or even usually, occur linearly or in isolation—rather, we discuss each one separately for the sake of simplicity only (Figure 11.4). We examine what might be involved in each type of thought de-automatization, and present evidence from both behavioral research and functional neuroimaging that supports the capacity of meditation training and hypnotic suggestion to facilitate this increased cognitive and emotional flexibility.

Furthermore, any of these de-automatization processes (Table 11.2; Figure 11.4) might conceivably be applied at any of the three putative stages of spontaneous thought (see Table 11.1), but throughout this chapter, we discuss them largely in relation to the *elaboration* stage, both for the sake of simplicity and also because the bulk of the cognitive neuroscientific evidence bears on the elaborative stage. In theory, however, de-automatization processes might be just as relevant for altering the frequency and type of thoughts arising during the *generation* stage, for instance by biasing the very content of thought—or in altering one's habitual judgments of one's own thoughts during the *evaluation* stage. A more detailed discussion of these possibilities awaits future theoretical work.

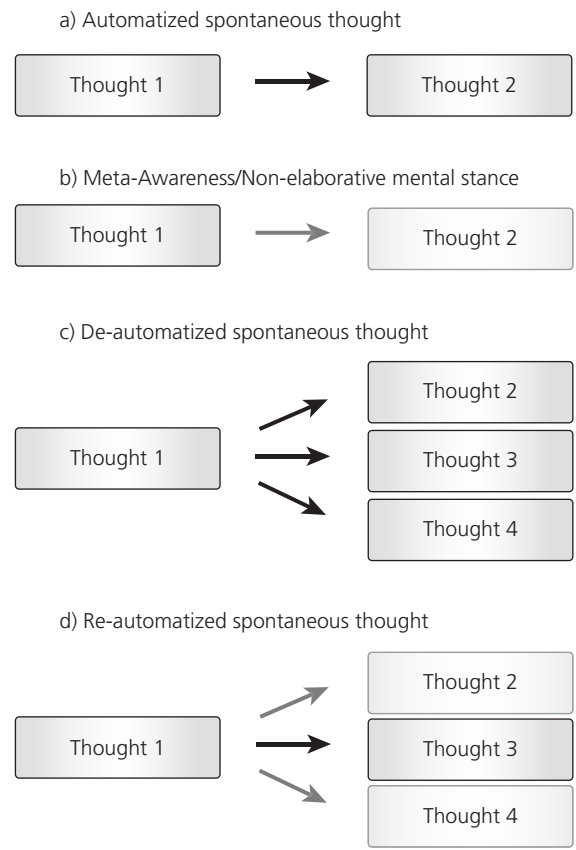

**Fig. 11.4** Schematic of automatized and more flexible patterns of spontaneous thought.

A schematic illustration of typical automatized spontaneous thought processes, and various ways in which cognition and emotion might become more flexible and less habitual with meditation practice or hypnotherapy; see also Table 11.2. **(a)** Typical thought is highly automatized, in that once a given *Thought 1* has arisen (either spontaneously, seemingly without reason; or perhaps dependent on a particular stimulus or a prior thought drifting below the threshold of awareness), it tends, strongly, to lead to a particular *Thought 2* (indicated by a strong black arrow). **(b)** With the non-elaborative mental stance adopted, for instance, by mindfulness meditation practitioners, thoughts are less likely to chain together into streams; the chance that *Thought 1* leads automatically to any other *Thought 2* is reduced (indicated by the "noisy," speckled arrow). **(c)** With de-automatization, thoughts are not necessarily any less likely to chain into streams, but the automaticity of these streams has been reduced: the stream of thought is now more flexible (i.e., characterized by less rigidity and more variability). A given *Thought 1* is now likely to lead to any number of other thoughts. **(d)** With the re-automatization of thought, a particular habitual stream of thought is chosen and cultivated. With practice, training, or suggestion, a given *Thought 1* is now more likely to lead to *Thought 3*, but not other thoughts (e.g., *Thought 2* and *Thought 4*).

# Reducing the chaining of spontaneous thoughts: a non-elaborative mental stance

Perhaps the simplest way (conceptually at least, but rarely so in practice) of changing the stream of spontaneous thought is to simply reduce the flow altogether. We refer to this reduction of the "chaining" of individual thoughts into an ongoing stream as a "non-elaborative" mental stance. A general goal of "quieting the mind" in this way is a central part of many meditation practices (e.g., Gunaratana, 2011; Iyengar, 2005).

There are numerous strategies for reducing the chaining of thoughts into elaborate and engrossing streams of fantasy, a major one being an increased focus on the sensory aspects of thoughts and perceptions (Fox et al., 2012; Kerr, Sacchet, Lazar, Moore, & Jones, 2013)—as opposed to allowing the mind to elaborate upon them with further thoughts about how they are related to one's self (e.g., Farb et al., 2007). For instance, one study of novices who underwent eight weeks of meditation training (Farb et al., 2007) subsequently presented them with various words (such as "cowardly," "envious," "cheerful," and "industrious") selected to spark self-related thinking about one's personality and behavior. In the first condition, "narrative" focus, participants were asked to dwell on the ideas evoked by these words: to think about whether and how each word related to their personality, what the words meant to them, and so on—in short, to engage in an elaborated stream of thoughts and judgments. In a second condition, "experiential" focus, they were asked to instead focus solely on whatever experience was evoked by the words in the present moment, whether it be thoughts, body sensations, or other mental events. Participants were asked to return their attention to the present moment if their thoughts started to run away with them—that is, to adopt a non-elaborative mental stance (Farb et al., 2007).

Compared to a wait-list control group with no meditation training, the meditation practitioners exhibited greater brain activity in several somatosensory and interoceptive brain regions during "experiential" versus "narrative" focus—and a corresponding decrease of activity in midline brain regions implicated in mind wandering (Figure 11.5 and Plate 6; Farb et al., 2007). Although this study did not directly address whether fewer thoughts were experienced when adopting a non-elaborative mental stance, it provided intriguing neural evidence that this might be the case (Figure 11.5).

Accumulating evidence supports the idea that mindfulness meditation can reduce the flow of thoughts: one recent study that involved two weeks of training in mindfulness meditation showed reduced self-reported rates of mind wandering in the mindfulness practitioners as compared to the control group (Mrazek, Franklin, Phillips, Baird, & Schooler, 2013). Another study instead examined long-term, expert meditation practitioners as compared to meditation-naïve control subjects (Brewer et al., 2011). Meditators self-reported less mind wandering during several forms of meditation practice and, concurrently, exhibited reduced brain activity in the same set of midline brain regions implicated in mind wandering as the aforementioned study (Farb et al., 2007).

Similar to certain forms of meditation, typical hypnotic induction procedures appear to quiet mental chatter, increase absorption, and alter activity within default mode regions associated with mind wandering. In one recent study, researchers induced hypnosis in an fMRI scanner and found that highly suggestible individuals showed reduced activity in classical default network regions including the right anterior cingulate gyrus, cortical midline structures of the left medial frontal gyrus, bilateral posterior cingulate cortices, and bilateral parahippocampal gyri (Deeley et al., 2012). These de-activations corresponded to self-reported increases in hypnotic depth and reductions in analytic thinking and mental chatter. An independent group showed similar dampening in default mode structures during hypnosis, albeit mostly restricted to the frontal components of this network (McGeown, Mazzoni, Venneri, & Kirsch, 2009). Furthermore, the degree of default mode dampening during the hypnotic induction predicted responsiveness to a subsequent hypnotic suggestion (Mazzoni, Venneri, McGeown, & Kirsch, 2013). Collectively, these data suggest that reductions in discursive thought and associated default mode activity may comprise a hallmark of hypnosis.

Although the aforementioned examples all suggest an *overall reduction* in the quantity of spontaneous thoughts with meditation and hypnosis, these experiments can only *suggest* reduced

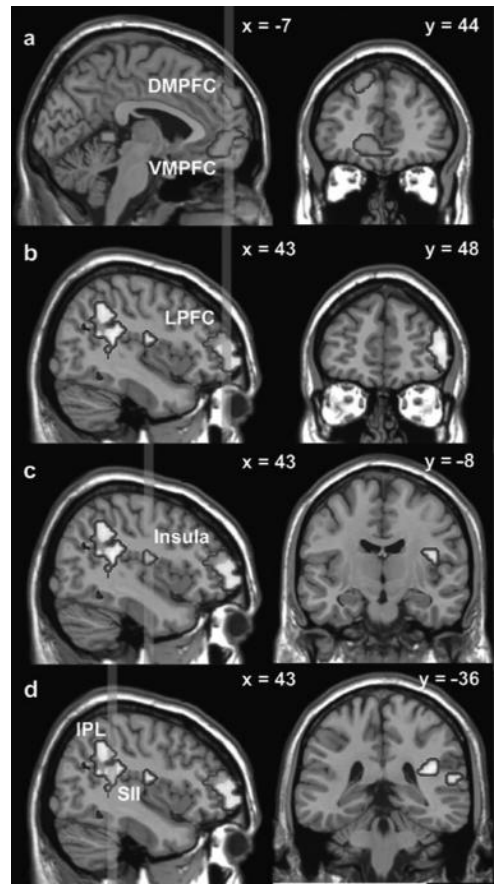

**Fig. 11.5** Neural correlates of a non-elaborative mental stance after brief meditation training (see also Plate 6).

After mindfulness meditation training, subjects were presented with words (e.g., "cowardly," "envious," "cheerful," "industrious") selected to spark self-related thinking about one's personality and behavior. When subjects were instructed to focus simply on their present- moment experience and do their best not to elaborate on the thoughts sparked by these words, those who had undergone meditation training showed markedly different brain activity compared to controls. Meditators showed reduced brain activity in medial prefrontal cortex areas implicated in self-referential processing **(a)**, and simultaneous increases of activity in several areas related to processing body image and interoceptive sensations, including the insula **(c)**, inferior parietal lobule **(d)**, and secondary somatosensory cortices **(d)**. Increased activity was also apparent in lateral prefrontal cortex areas **(b)** implicated in metacognition and executive control, suggesting a heightened awareness of, and perhaps control over, spontaneously arising thoughts. These data support the notion that the adoption of a non-elaborative, present-centered mental stance, grounded in the body, can reduce our habitual elaboration of thoughts.

Reproduced from Norman A. S. Farb, Zindel V. Segal, Helen Mayberg, Jim Bean, Deborah McKeon, Zainab Fatima, and Adam K. Anderson, Attending to the present: mindfulness meditation reveals distinct neural modes of self-reference, *Social Cognitive and Affective Neuroscience*, 2 (4), pp. 313–322, Figure 3, doi: 10.1093/scan/nsm030 (c) 2007, Oxford University Press.

chaining of thoughts. A recent study from our group, however, has directly investigated chaining of thoughts in expert meditation practitioners, with the aim of differentiating the neural correlates of a single thought that arises and passes away, from one which develops into an ongoing stream of thought (Ellamil et al., in preparation).

To address this question, we had participants in the fMRI scanner indicate the arising of a thought by pressing a button—but beyond this, they also indicated whether a *single* thought had arisen and then passed away, or whether the first thought had been elaborated into a *chain* of

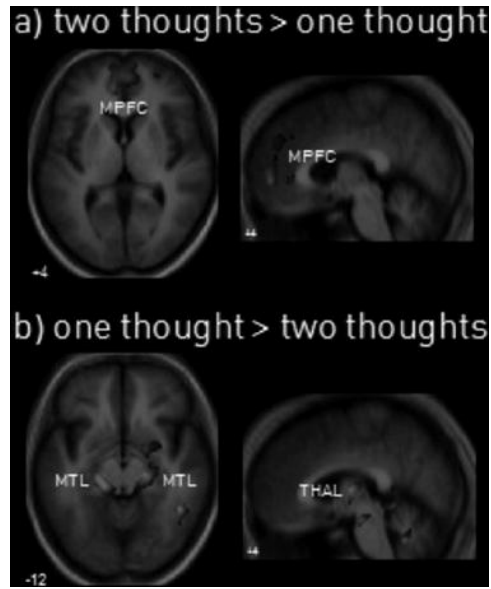

**Fig. 11.6** Neural correlates of single transient thoughts versus further elaboration of thoughts (see also Plate 7).

Participants highly experienced in introspection (long-term meditation practitioners) engaged in a simple, ostensibly thought-free form of meditation and then indicated, by pressing a button, whenever a thought arose in their minds. Additionally, they indicated whether only a single thought had arisen and then passed away, or whether the generation of a given thought was followed by subsequent chaining of this thought with others into an elaborated stream of two or more thoughts. Comparing brain activation across these two conditions, our group found that the generation of a single thought (panel b) is preferentially dependent on memory centers of the medial temporal lobe (MTL) and sensory relay areas in the thalamus (THAL). In contrast, chaining of thoughts (panel a) into an elaborated stream more powerfully recruited medial prefrontal cortex (MPFC) areas strongly implicated in mind wandering and related forms of spontaneous thought.

Figure reproduced with permission from Ellamil et al. (in preparation).

two or more thoughts. We then compared neural activity associated with single versus chained thoughts and found that single spontaneous thoughts seem to arise dependent on recruitment of memory centers of the medial temporal lobe and also to recruit sensory relay centers of the thalamus (Figure 11.6b). However, similar to the results of the studies already mentioned (Brewer et al., 2011; Farb et al., 2007), the elaboration of individual thoughts into connected streams of thinking more strongly recruited brain areas such as the medial prefrontal cortex (Figure 11.6a).

Collectively, these studies (Brewer et al., 2011; Ellamil et al., in preparation; Farb et al., 2007) suggest that medial prefrontal cortex areas are involved in the automatized elaboration of thought streams, and that paying increased attention to present-moment experience (as taught in various meditation traditions, or as induced by hypnotic suggestion) can dampen activity in these areas, reduce the chaining of thoughts, and upregulate activity in brain areas related to exteroceptive and interoceptive body sensation.

## De-automatization of spontaneous thinking: broadening the stream of thought

In contrast to a non-elaborative mental stance, which aims at reducing the chaining and quantity of thoughts (Figure 11.4b), the goal of de-automatization is to derail habitual patterns and thus broaden the thought stream. Instead of a narrow, prescribed course, with de-automatization, a given thought or stimulus may lead to any number of subsequent thoughts or emotional experiences (Figure 11.4c). With de-automatization, spontaneous thinking is not necessarily reduced, but greater flexibility and variety is now present (Figure 11.4c).

## Hypnosis overrides automatic cognitive processes

In recent years, a growing body of cognitive neuroscience evidence has demonstrated that suggestion can swiftly override a wide range of deeply ingrained processes. Perhaps one of the most striking and well-known examples of de-automatization is a series of studies showing that suggestion can reduce, or in some cases even eliminate, the classic "Stroop" effect (Lifshitz, Aubert Bonn, Fischer, Kashem, & Raz, 2013; Raz, Fan, & Posner, 2005; Raz, Kirsch, Pollard, & Nitkin-Kaner, 2006; Raz, Moreno-Íniguez, Martin, & Zhu, 2007; Raz, Shapiro, Fan, & Posner, 2002). In the Stroop paradigm, incongruent trials—in which participants are asked to report the color of a word displayed in an incompatible font color (e.g., the word "blue" displayed in red font)—lead to conflict between two automatized cognitive reactions (i.e., reading basic words and perceiving color). Thus, incongruent trials are strongly associated with slower reaction times and greater activity in neural regions involved in cognitive conflict, such as the dorsal anterior cingulate cortex (Figure 11.7 and

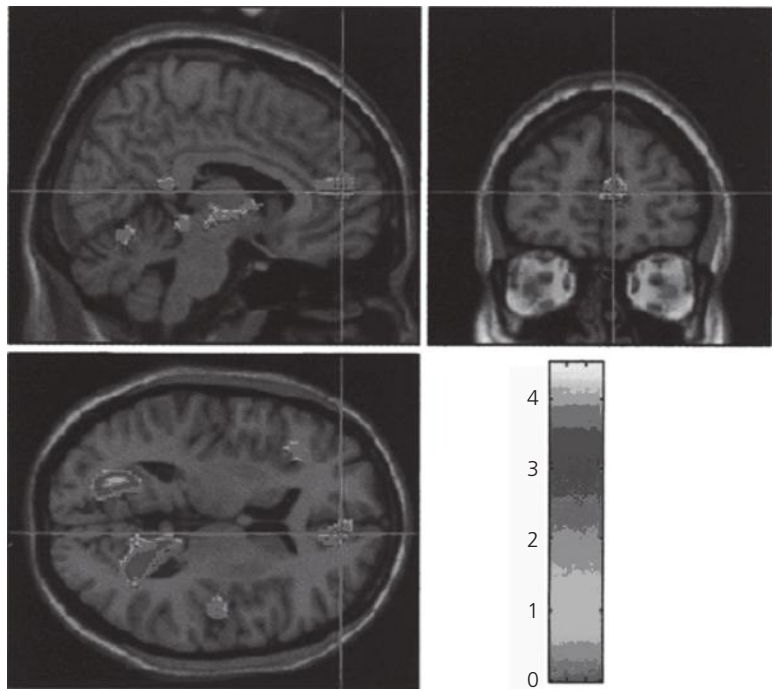

**Fig. 11.7** Brain activations during Stroop task conflict are reduced by hypnotic suggestion (see also Plate 8).

Brain regions showing *less* activation, after hypnotic suggestion, during incongruent versus congruent trials on the Stroop task. Note, in particular, the reduced activity in the anterior cingulate cortex (under crosshairs), a region strongly implicated in conflict processing and resolution. Alongside behavioral results indicating less conflict, and equal performance, in the hypnotized condition, these fMRI findings suggest that hypnotic suggestion can reduce automatized cognitive reactions to perceptual stimuli.

Adapted from Amir Raz, Jin Fan, and Michael I. Posner, Hypnotic suggestion reduces conflict in the human brain, *Proceedings of the National Academy of Sciences of the United States of America*, 102 (28), pp. 9978–9983, Figure 1a, doi: 10.1073/pnas.0503064102, Copyright © 2005, The National Academy of Sciences.

Plate 8; Raz et al., 2005). Yet, Raz et al. (2002) showed that highly suggestible individuals can attenuate, or in some cases even abolish, this Stroop interference effect following a rapid and straightforward suggestion to perceive the stimulus words as meaningless symbols of a foreign language. Multiple studies from independent groups around the world have since replicated these initial findings (e.g., Augustinova & Ferrand, 2012; Casiglia et al., 2010; Parris, Dienes, Bate, & Gothard, 2013; Parris, Dienes, & Hodgson, 2012, 2013; Raz & Campbell, 2011; Raz et al., 2005, 2006, 2007).

Results from a seminal functional neuroimaging study (Raz et al., 2005) support the collective behavioral results. In the absence of suggestion, strong activations were seen throughout the brain in response to incongruent versus congruent Stroop trials, as might be expected by the conflicting automatic processes inherent to incongruent trials. Most notable during suggestion-free incongruent trials was significantly increased activation in the dorsal anterior cingulate cortex, known to have a critical role in both monitoring and resolving cognitive conflict (Figure 11.7; Raz et al., 2005). Following suggestion, however, this conflict was reduced both behaviorally and neurally: reaction time and accuracy were now comparable on congruent and incongruent trials, and significantly less brain activity occurred within the anterior cingulate cortex (Figure 11.7; Raz et al., 2005). Interested readers may wish to consult Lifshitz et al. (2013) for a fuller exposition of the putative mechanisms and interpretational issues surrounding these findings (see also Chapter 16).

A number of recent studies have extended the prospect of hypnotic de-automatization beyond Stroop interference to other deeply ingrained processes. In the realm of visual attention, suggestion improved performance on two classic paradigms probing involuntary response conflict: the Flanker (Iani, Ricci, Gherri, & Rubichi, 2006) and Simon (Iani, Ricci, Baroni, & Rubichi, 2009) tasks. Another study recently demonstrated that hypnosis could derail cross-modal perception in the classic McGurk effect—an auditory illusion crafted by presenting visual and auditory streams that are incongruent, demonstrating the influence of visual facial movements on auditory speech percepts (McGurk & MacDonald, 1976). So robust is the McGurk effect that people are typically unable to avoid the illusion even if they are aware of the audiovisual discrepancy (McGurk & MacDonald, 1976) and regardless of practice (Summerfield & McGrath, 1984). Yet, a simple suggestion for increased auditory acuity greatly reduced illusory speech percepts and improved correct audio identifications on the McGurk task (Déry, Campbell, Lifshitz, & Raz, 2014). Along similar lines, researchers were able to override cross-modal perceptual integration in a single, highly hypnotically suggestible face-color synesthete, eliciting concomitant alterations in her event-related potential (ERP) profile (Terhune, Cardeña, & Lindgren, 2010).

Thus, the potential for using suggestion to de-constrain habitual cognitive patterns seems to generalize beyond the Stroop effect and offers intriguing prospects for further cognitive and applied investigations, including in the realms of spontaneous thought and its disorders. For example, in Chapter 23, Lynn et al. offer an intriguing account of how suggestion-based de-automatization can translate to meaningful clinical outcomes related to craving and addiction.

## Improving implicit intergroup attitudes

Outside of the laboratory, is there a possibility of de-automatizing spontaneous thought processes with direct relevance to everyday life? One such form of detrimental spontaneous thinking is the formation and maintenance of implicit intergroup attitudes. Implicit judgments against stigmatized social groups are often automatically activated without conscious awareness, and typically measured with response-latency techniques such as the Implicit Association Test (IAT; Greenwald, Poehlman, Uhlmann, & Banaji, 2009) to circumvent limitations of introspective self-reports.

   

While implicit intergroup attitudes are highly resistant to change due to their automatic and unconscious nature (Devine, 1989), recent findings suggest that de-automatization of such biases is possible through the practice of meditation (Kang, Gray, & Dovidio, 2013). In one study, loving-kindness meditation—a concentration practice that aims to establish a deep sense of positive interconnectedness to others (Salzberg, 2004)—improved automatic intergroup orientations. Specifically, participants who completed six weeks of loving-kindness meditation training showed reduced implicit bias toward members of two socially stigmatized outgroups, Black and homeless people, as measured by IATs (Kang, Gray, & Dovidio, 2013). Importantly, the reduction of implicit bias was not present in the closely matched active controls who attended a six-week loving-kindness group discussion course that aimed to understand and share ideas of loving-kindness and compassion in the absence of any actual meditation training. This suggests that mere conceptual understanding of compassion may be insufficient for the de-automatization of such deeply ingrained attitudes; actual practices such as meditation may be necessary.

## Reducing the affective-elaborative component of pain

Yet another form of automatized thought that many would presumably like to reduce is the emotional and cognitive distress and yearning for cessation that tends to accompany physical discomfort and pain. (For a much more in-depth take on pain, hypnosis, and meditation, see Chapter 21). Aside from the purely sensory aspects of pain, nociception is ubiquitously attended by elaborative thoughts that relate the pain to the self, and by affective experiences involving distress and various other negative emotional valuations of the sensory experience (Grant, 2014; Grant, Courtemanche, & Rainville, 2011; Rainville, Duncan, Price, Carrier, & Bushnell, 1997).

Beyond these subsequent *elaborations* of painful stimuli, even the mere *expectation* or anticipation of pain can influence the subsequent amplitude of an actual pain experience. For instance, in one study, the mere expectation of a painful stimulus appeared to amplify the actual experience of unpleasantness in response to a normally innocuous stimulus, as indexed by increased brain responses within areas implicated in pain processing (Sawamoto et al., 2000).

These cognitive-emotional elaborations appear to be a habitual, spontaneous response to a painful sensation, rather than an intrinsic aspect of nociception itself (Grant, 2014). They appear to be dissociable from, and temporally subsequent to, the purely sensory aspects of pain—and what is more, they may contribute significantly to the subjectively experienced unpleasantness of nociceptive experience (Rainville et al., 1997). As physical pain and discomfort are an unavoidable part of everyday life, and a severe burden in clinical disorders such as neuropathic pain (Woolf & Mannion, 1999) and chronic migraine (Olesen et al., 2006), an intervention that reduces our habitual tendency to engage in negative cognitive-emotional valuation of nociceptive experience is of obvious benefit to healthy people, as well as those suffering from certain clinical conditions.

Numerous studies over the past few decades have shown that mindfulness meditation indeed appears to be effective for the relief of various forms of chronic pain (e.g., Kabat-Zinn, 1982; Kabat-Zinn, Lipworth, & Burney, 1985; Morone, Greco, & Weiner, 2008)—but what might be the cognitive-neural mechanisms of this change? A recent study using fMRI explored whether mindfulness meditation might reduce the habitual affective-elaborative component of pain, by exploring differences in brain activity between long-term Zen practitioners and meditation-naïve controls (Grant et al., 2011). In line with earlier studies showing that the same sample of Zen practitioners exhibits higher self-reported pain thresholds (Grant & Rainville, 2009) and increased cortical thickness in sensory pain areas (Grant, Courtemanche, Duerden, Duncan, & Rainville, 2010),

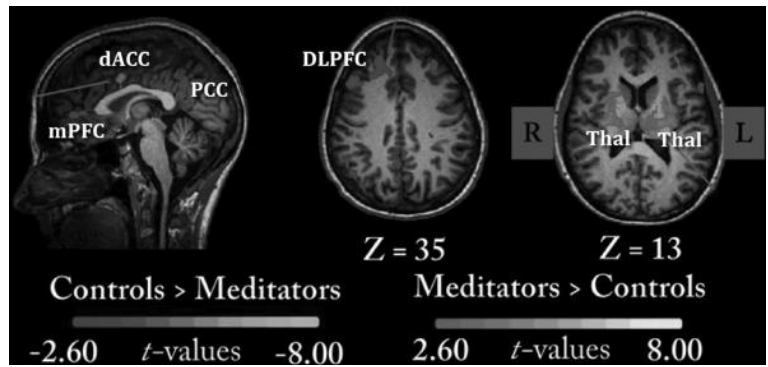

**Fig. 11.8** Meditators show less affective-elaborative brain activation in response to pain (see also Plate 9).

In response to a painful thermal stimulus, long-term meditation practitioners showed greater brain activations in primary sensory pain areas, including the dorsal anterior cingulate cortex (dACC) and bilateral thalamus (Thal). In contrast, meditation-naïve controls showed greater activation in brain areas strongly implicated in elaborative and self-referential thought, including the medial prefrontal cortex (mPFC), posterior cingulate cortex (PCC), and dorsolateral prefrontal cortex (DLPFC). Z values indicate vertical position in stereotactic Talairach space.

Adapted from *Pain* 152 (1), Joshua A. Grant, Jérôme Courtemanche, and Pierre Rainville, A non-elaborative mental stance and decoupling of executive and pain-related cortices predicts low pain sensitivity in Zen meditators, pp. 150–156, doi:10.1016/j.pain.2010.10.006, © 2011, International Association for the Study of Pain.

the researchers found that pain elicited greater brain activations in sensory pain areas in the meditators, including the thalamus, insula, and dorsal anterior cingulate cortex—suggesting that the meditators elaborated less on their pain, and instead focused more on its immediate sensory components (Figure 11.8 and Plate 9; Grant et al., 2011).

Further support for this interpretation came from the finding that functional connectivity (temporal correlation of activity patterns—Buckner, Krienen, & Yeo, 2013; Van Dijk et al., 2010) among the sensory pain areas was strengthened in the long-term meditation practitioners. On the other hand, functional connectivity between primary sensory pain areas and secondary affective-elaborative areas was increased in controls and decreased in long-term meditation practitioners (Grant et al., 2011), again supporting the idea that novices continued to elaborate upon their pain with further emotional and cognitive associations, whereas the meditators tended to remain focused on its purely sensory aspects. Intriguingly, the degree of this decoupling in meditation practitioners significantly predicted their lower pain sensitivity—offering perhaps the most convincing evidence for the aforementioned interpretations of the neuroimaging results.

Although the aforementioned study was not able to directly address the question of whether reduced cognitive-affective elaboration of the sensory pain experience was the *direct* cause of the observed lowered pain sensitivity and cortical decoupling, the results are nonetheless suggestive. (For a detailed analysis of the mechanisms underlying analgesia via mindfulness and hypnosis, see Chapter 21.) Note, too, that this example of reducing the affective-elaborative component of pain might also be considered an example of a "non-elaborative mental stance" (see the section "Reducing the chaining of spontaneous thought: a non-elaborative mental stance"). However, as noted in the section "Recognizing and combating the automaticity of spontaneous thought processes: the benefits of meta-awareness," we distinguish each sub-form of de-automatization (Table 11.2; Figure 11.4) largely for the sake of clarity—and not because of any strong demarcations between these various means of increasing cognitive-emotional flexibility.

## Re-automatization of spontaneous thought processes

The philosopher P. D. Ouspensky aptly described the cognitive process we refer to here as *re-automatization*:

> We do not realize what enormous power lies in thinking . . . The power lies in the fact that, if we always think rightly about certain things, we can make it permanent—it grows into a permanent attitude . . . If you start from right thinking, then after some time you will educate in yourself the capacity for a different reaction. (Ouspensky, 1957, pp. 76–77)

As desirable as conscious, flexible thought patterns may be, automatization (as already noted) is beneficial in at least two respects: automatic behaviors and mental processes require fewer attentional resources and, relatedly, are faster (Moors & De Houwer, 2006). Therefore, altering patterns of spontaneous thought and emotion should not be considered solely a "negative" or reductive process: there is room, too, for constructive action in the creation of healthy, optimal habitual thinking that then effortlessly promotes one's well-being (Figure 11.4d).

This notion is familiar in meditation traditions: a practitioner engaging in loving-kindness meditation, for instance, might seek to change their usual habit of responding with irritation or anger to personally offensive stimuli, to a new pattern of responding with compassion and understanding instead. Similarly, practitioners of compassion meditation train themselves to respond to signs of others' pain and distress with compassion and a genuine desire to alleviate their suffering, instead of prior habitual reactive patterns of indifference or pity. An fMRI study (Lutz, Brefczynski-Lewis, Johnstone, & Davidson, 2008) in which participants were exposed to emotionally negative human vocalizations (i.e., sounds indicative of human pain and distress) found some evidence supporting the view that meditation might effect such re-automatization. Compared to control participants with no meditation experience, expert compassion meditation practitioners showed increased blood oxygen level dependent (BOLD) signal in the superior temporal sulcus, amygdala, and temporoparietal junction. These regions have been linked to auditory and emotional processing, as well as "theory of mind" (imagining the beliefs and intentions of others) and empathy (Völlm et al., 2006), suggesting, to the authors, not only enhanced processing of the sounds themselves but also increased compassionate responses in the meditators (Lutz et al., 2008). The long-term compassion meditation practitioners seemed to have successfully re-automatized their reactions to negative human vocalizations, such that their default response was no longer fear or worry, but rather compassionate concern.

Subsequent work supports the idea that brief meditation training can re-orient people from habitually selfish, to increasingly altruistic, behavior (Condon, Desbordes, Miller, & DeSteno, 2013; Weng et al., 2013), and might do so by altering neural function in the nucleus accumbens, which plays a major role in the neurochemical dopamine signaling that gives rise to our subjective sense of what feels fulfilling and rewarding (Weng et al., 2013).

Another meditation study suggestive of re-automatization divided participants into three groups: novice meditation practitioners, expert practitioners with approximately 19,000 hours of experience, and extremely advanced expert practitioners with an average of approximately 44,000 hours of experience (Brefczynski-Lewis, Lutz, Schaefer, Levinson, & Davidson, 2007). During a demanding, sustained attention task (typically very challenging for beginners), brain activity in attention networks showed an inverted U shape across the three groups: activity was highest in the expert meditators, and lower in both novices and the extremely advanced experts (Brefczynski-Lewis et al., 2007). This finding suggested, to the authors, that the expert meditators had learned to increase recruitment of these regions to consciously sustain attention in an effortful manner (which the novices could not do), but that in extremely advanced practitioners, this process had become relatively automatic and effortless, resulting in brain activity similar to that seen in novices (Brefczynski-Lewis et al., 2007). As noted in the section "The automaticity of spontaneous thought: advantages and disadvantages," "automaticity" denotes a process that has become essentially effortless with either continued practice or repeated exposure. This inverted U pattern of brain activation suggests that concentration meditation is initially difficult or almost impossible for those without training; with increased training, an effortful process of learning takes place, as suggested by greater brain activation; and, finally, sustained attention becomes ever more automatized with continued practice, resulting in minimal brain activation, comparable to that of novices.

In the realm of hypnosis, burgeoning research indicates that suggestion may be capable of shifting cognitive processes from effortful to automatic without extensive practice (Lifshitz et al., 2013). Preliminary support for this prospect comes from a study showing that a suggestion engendered digit-color synesthesia effects in highly hypnotizables by promoting perceptual integration across sensory modalities (Kadosh, Henik, Catena, Walsh, & Fuentes, 2009). More recently, two pilot experiments explored the prospects of rendering an effortful task easier in the realm of visual attention (Lifshitz et al., 2013). The first experiment involved the masked-diamond paradigm, in which participants identify the direction of moving geometric figures (e.g., clockwise, counter-clockwise) with invisible apexes. When a visual mask occludes the invisible apexes, motion detection is immediate and effortless; without the occluding masks, however, determining the direction of motion is extremely difficult for most people (see http://razlab.mcgill.ca/demomotrak.html). Pilot results show that highly suggestible, but not less suggestible, individuals were able to perform the task with remarkable accuracy following a suggestion to visualize the occluding masks, indicating that they could shift the difficult, almost intractable, task into the realm of automaticity. The second pilot experiment involved a classic visual search task, in which participants scan a display for a target item among distractors. Among highly suggestible individuals, a suggestion to see the target pop out effortlessly from the distractors significantly improved the efficiency of search.

These early findings support the prospect of using suggestion to render processes more automatic without extensive exposure or practice. Indeed, clinical hypnosis practitioners have been leveraging this potential for decades in helping patients replace undesired mental activities (e.g., negative ruminations in depression) with more wholesome cognitive processes (e.g., fostering optimistic outlooks) (Yapko, 2013).

Re-automatization, then, is similar to everyday automatized thought, in that it involves automatic, "habitual" routines. The main difference is that thought is now re-oriented toward goals and thinking patterns one approves of or desires. The aim, with training, is to more effortlessly think in ways that one deems right, good, or desirable. In many cases, it would seem ideal that good habits and patterns of thought would also become automatized and, therefore, relatively effortless and non-resource-draining.

## Discussion and future directions

In this chapter, we have outlined numerous ways in which our patterns of thinking and feeling might be made more flexible via meditation training and hypnotic suggestion (Table 11.1; Figure 11.4). We have also touched on the results of some seminal studies that point toward a broad understanding of the functional neural correlates of these changes, as measured with neuroimaging tools such as fMRI. In this final section, we speculate on the neuroanatomical and molecular neurobiological changes that might underlie the functional neural plasticity associated with cognitive-emotional flexibility.

## Do stable neuroanatomical changes accompany increased cognitive-emotional flexibility and functional neural plasticity?

Throughout this chapter, we have discussed cognitive-emotional flexibility, and functional neural plasticity, related to de-automatization, but we have not addressed potential neuro*anatomical* changes that might accompany these behavioral and functional differences. The cognitive and functional neural changes discussed throughout this chapter imply neuroanatomical alterations however—perhaps not just at the level of synapses, but potentially on a scale observable with non-invasive neuroimaging methods (Zatorre, Fields, & Johansen-Berg, 2012). In particular, it seems plausible that long-term, persistent engagement in hypnosis or meditation training might lead to structural brain plasticity subserving the more readily observable changes in behavior and brain activity. The study of these morphological brain differences in humans with non-invasive neuro-imaging methods is known as "morphometric" neuroimaging, and aims to characterize various structural aspects of the brain's gray and white matter (see, e.g., Draganski & May, 2008; May & Gaser, 2006; Zatorre et al., 2012).

To our knowledge, very little work has addressed the relationship between neuroanatomical structure and hypnosis (though for two seminal studies exploring neuroanatomical correlates of high hypnotic suggestibility, see Horton, Crawford, Harrington, & Downs, 2004; Huber, Lui, Duzzi, Pagnoni, & Porro, 2014). On the other hand, more than 20 studies have now investigated meditation practitioners using morphometric neuroimaging methods (for some seminal studies, see Grant et al., 2010; Holzel et al., 2008; Lazar et al., 2005; Pagnoni & Cekic, 2007).

Recently, some of us (Fox, Nijeboer, et al., 2014) undertook a comprehensive review and meta-analysis of these morphometric studies of meditation practitioners (Figure 11.9 and Plate 10). Among numerous intriguing differences in gray and white matter, some consistent meta-analytic clusters are of particular relevance to the preceding discussion of de-automatization. Consistent neuroanatomical differences were observed in the rostrolateral prefrontal cortex (Brodmann area 10), for instance (Figure 11.9, left panel), which is strongly implicated in meta-cognitive awareness and accuracy (Christoff & Gabrieli, 2000; Christoff, Ream, Geddes, & Gabrieli, 2003; Fleming & Dolan, 2012; Fleming, Dolan, & Frith, 2012; Fleming, Weil, Nagy, Dolan, & Rees, 2010; McCaig, Dixon, Keramatian, Liu, & Christoff, 2011). As we have already noted, meta-awareness of the automaticity of spontaneous thought patterns seems a prerequisite to any enduring change of these patterns (see the section "Recognizing and combating the automaticity of spontaneous thought processes: the benefits of meta-awareness"); structural plasticity in what is arguably the key meta-cognitive brain region might therefore support these changes.

Also of interest are meta-analytic differences in the insular cortex (Figure 11.9, left panel; Fox, Nijeboer, et al., 2014). Although the insula has been implicated in a wide variety of cognitive and emotional processes (Menon & Uddin, 2010; Singer, Critchley, & Preuschoff, 2009), its role in interoception (awareness of internal bodily sensations) is particularly prominent (Craig, 2004, 2009; Critchley et al., 2004). As discussed in the section "Reducing the chaining of spontaneous thought," a non-elaborative mental stance often involves increased focus on the immediate, present-centered sensations from within the body as a means of avoiding becoming ensnared in automatized streams of thought (Farb et al., 2007; Fox et al., 2012; Kerr et al., 2013). Structural differences in the insula might therefore play a role in the increased awareness of the body and present-moment experience cultivated by meditation practitioners (Farb, Segal, & Anderson, 2013a, 2013b; Farb et al., 2007; Fox et al., 2012; Kerr et al., 2013).

A further area of interest is the medial temporal lobe, particularly the hippocampus and par-ahippocampal cortex (Figure 11.9, right panel), which appear to exhibit altered structure in

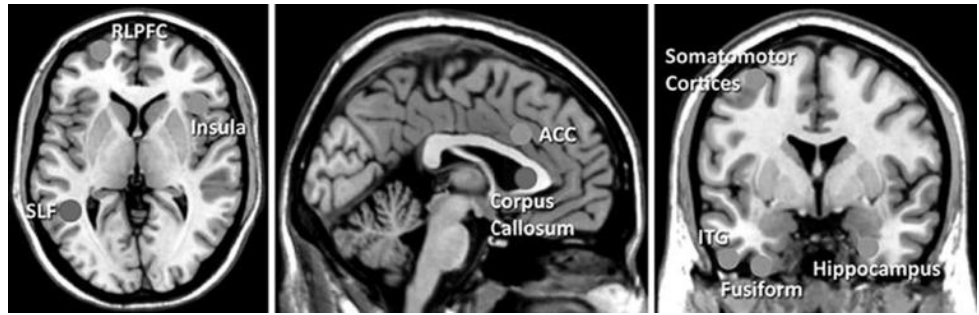

**Fig. 11.9** Brain regions showing consistent structural differences in meditation practitioners (see also Plate 10).

Brain regions which show significant structural differences in meditation practitioners versus meditation-naïve controls, as measured via various morphometric neuroimaging techniques. Schematic results based on a meta-analysis of all morphometric neuroimaging studies of meditation practitioners. ACC: anterior cingulate cortex; ITG: inferior temporal gyrus; SLF: superior longitudinal fasciculus; RLPFC: rostrolateral prefrontal cortex.

meditation practitioners (Fox, Nijeboer, et al., 2014). The well-established role of the medial temporal lobe in the consolidation of memory (Squire, 2004, 2009; Squire, Stark, & Clark, 2004) has more recently been complemented by recognition of its possible role in the *re*consolidation of memories that have been reactivated (Finnie & Nader, 2012; Nader, Schafe, & Le Doux, 2000; Schwabe, Nader, Wolf, Beaudry, & Pruessner, 2012). Structural differences in medial temporal lobe structures may therefore be related to the reorganization of memories themselves, and of one's habitual reactions to spontaneously arising recollections. A more detailed discussion of this possibility follows.

## What is the molecular neurobiological basis of de-automatization?

Even with a basic model of various forms of de-automatization (Figure 11.4), and identification of some potential neuroanatomical correlates, the deeper mystery remains of what cellular and molecular neurobiological mechanisms might subserve the cognitive-emotional flexibility observable at a larger scale as functional and structural brain plasticity and cognitive de-automatization.

The emerging field of memory *reconsolidation* might speak to this issue. For decades, the general consensus was that after an initial period of consolidation, a memory was basically stable. A growing literature, however, supports the idea that *reactivated* memories again enter a labile state where their consolidation can be disturbed, and the memory thereby weakened (Duvarci & Nader, 2004; Duvarci, Nader, & LeDoux, 2005, 2008; Finnie & Nader, 2012; Nader et al., 2000; Schwabe et al., 2012). If memories are reactivated in some manner, a brief time window opens during which these memories can be tampered with, pharmacologically, and their subsequent consolidation reduced or potentially blocked entirely. We propose that if memories arise again spontaneously, during mind wandering for instance, they ought to enter a similarly labile state and be open to deconsolidation or reconsolidation at the molecular level.

A mechanistic account suggests that emotional enhancement of memory is one major route whereby such reconsolidation of memory might take place. As emotion (both positive and negative) enhances memory for many kinds of stimuli (Markovic, Anderson, & Todd, 2014), blocking the neurobiological pathways mediating this emotional modulation during the period of memory reactivation might prevent, or at least modulate, reconsolidation. Several studies, targeting a variety of possible neurobiological mechanisms (including mRNA (messenger ribonucleic acid) synthesis and protein synthesis) of emotionally modulated memory, have now shown such results, usually in relation to fear-related or otherwise negatively valenced memories (Duvarci & Nader, 2004; Duvarci et al., 2005; Duvarci, Nader, & LeDoux, 2008; Nader et al., 2000).

Many of these studies have been conducted in laboratory animals (rats in particular), inviting the question of how relevant such deconsolidation of fear memory might be for our own species. One intriguing study directly addresses this concern, however, by employing a pharmacological intervention in human subjects. By administering the drug propranolol, a β-adrenergic receptor antagonist, Kindt, Soeter, and Vervliet (2009) were able to demonstrate analogous fear memory deconsolidation effects in humans (Kindt et al., 2009). Another study followed up on these results by performing a similar experiment alongside acquisition of fMRI data: Schwabe et al. (2012) administered either propranolol, or a placebo, during the reactivation of previously acquired emotional or neutral material. Their findings supported the earlier propranolol study (Kindt et al., 2009) in showing impaired reconsolidation of emotional material in the propranolol, but not placebo, group (Schwabe et al., 2012). FMRI showed corresponding differential neural activity, during both reactivation and testing of the emotional memories, in the amygdala and hippocampus—two of the key brain structures involved in memory reconsolidation (Schwabe et al., 2012).

Due to the strong reliance of spontaneous thought processes upon memories (Andrews-Hanna, Reidler, Huang, et al., 2010; Fox et al., 2013; Klinger, 2008), the tendency of memories to chain together into an associated stream of thought (e.g., Ellamil et al., in preparation; Figure 11.6 and Plate 7), and the ubiquity of emotion in spontaneous thoughts (Fox et al., 2013; Fox, Thompson, et al., 2014), these results suggest the intriguing possibility that non-pharmacological interventions like meditation and hypnotherapy might be able to reduce the occurrence, emotional intensity, or chaining of thoughts and memories. Damping one's emotional reactivity to spontaneously arising thoughts and moment-to-moment experience, for instance (Feldman, Greeson, & Senville, 2010), might therefore lead to similar "deconsolidation" events at the molecular neurobiological level, preventing older patterns of thought from being reconsolidated and perpetuated. These ideas remain highly speculative, however, and await detailed future work.

## Conclusion

The great neuroanatomist Santiago Ramón y Cajal once wrote, "The youthful brain is wonderfully pliable and, stimulated by the impulses of a strong will to do so, can greatly improve its organization by creating new associations between ideas and by refining the powers of judgment" (Ramon y Cajal, 2004, p. 23). We believe that the evidence reviewed here supports a similar conclusion for not-so-youthful adult brains as well. Under appropriate conditions and with effective techniques (for instance, training from a meditation instructor, or suggestion from a hypnotherapist), there appears to be an impressive degree of flexibility at both the cognitive and neural level.

## Acknowledgments

The authors thank Dr. Norman A. S. Farb, Dr. Amir Raz, Dr. Melissa Ellamil, Dr. Joshua A. Grant, and Dr. Matthew L. Dixon for their kind permission to reproduce portions of figures from their original research throughout this chapter. This work was supported, in part, by Natural Sciences and Engineering Research Council (NSERC) Vanier Canada Graduate Scholarships awarded to K. C. R. F and M. L., and research grants from the Canadian Institutes of Health Research (CIHR) and NSERC awarded to K. C.